\documentclass[journal,twocolumn]{IEEEtran}
\usepackage{cite}
\usepackage{bm}
\usepackage{algorithm}
\usepackage{algcompatible}
\usepackage{float}
\usepackage{array}
\usepackage{multirow}
\usepackage[cmex10]{amsmath}
\usepackage{graphicx}
\usepackage{url}
\usepackage{color}
\usepackage{amssymb}
\usepackage{amsfonts}
\usepackage{amsmath}
\usepackage{extarrows}
\usepackage{graphicx}
\usepackage{amsfonts}
\usepackage{caption}
\usepackage{algorithm}
\usepackage{algpseudocode}
\usepackage{indentfirst}
\usepackage{caption}
\usepackage{esint} 
\usepackage{graphicx}
\usepackage{graphics}
\usepackage{pgfplots}
\usepackage{tikz}
\usepackage{epsfig}
\usepackage{soul}
\usepackage{caption}
\usepackage{gensymb}
\def\bA{{\mathbf{A}}} \def\bB{{\mathbf{B}}} \def\bC{{\mathbf{C}}} \def\bD{{\mathbf{D}}} \def\bE{{\mathbf{E}}}
 \def\bG{{\mathbf{G}}} \def\bH{{\mathbf{H}}} \def\bI{{\mathbf{I}}} \def\bJ{{\mathbf{J}}}
   \def\bN{{\mathbf{N}}} 
\def\bP{{\mathbf{P}}} \def\bQ{{\mathbf{Q}}}   \def\bT{{\mathbf{T}}}
\def\bU{{\mathbf{U}}}  \def\bW{{\mathbf{W}}}  \def\bY{{\mathbf{Y}}}

\def\ba{{\mathbf{a}}}    
  \def\bh{{\mathbf{h}}}  
   \def\bn{{\mathbf{n}}} 
\def\bp{{\mathbf{p}}}    
   \def\bx{{\mathbf{x}}} \def\by{{\mathbf{y}}}

\begin{document}

\title{STAR-RIS-Enabled Simultaneous Indoor and Outdoor 3D Localization: Theoretical Analysis and Algorithmic Design

\author{Jiguang~He,~\IEEEmembership{Senior Member,~IEEE,} Aymen~Fakhreddine,~\IEEEmembership{Member,~IEEE,} and George C. Alexandropoulos,~\IEEEmembership{Senior Member,~IEEE}}
\thanks{J. He, A. Fakhreddine, and G. C. Alexandropoulos are with the Technology Innovation Institute, 9639 Masdar City, Abu Dhabi, United Arab Emirates. G. C. Alexandropoulos is also with the Department of Informatics and Telecommunications, National and Kapodistrian University of Athens, 15784 Athens, Greece. (e-mails: \{jiguang.he, aymen.fakhreddine\}@tii.ae, alexandg@di.uoa.gr).}}
 \maketitle

\begin{abstract}
Recent research and development interests deal with metasurfaces for wireless systems beyond their consideration as intelligent tunable reflectors. Among the latest proposals is the simultaneously transmitting (a.k.a. refracting) and reflecting reconfigurable intelligent surface (STAR-RIS) which intends to enable bidirectional indoor-to-outdoor, and vice versa, communications thanks to its additional refraction capability. This double functionality provides increased flexibility in concurrently satisfying the quality-of-service requirements of users located at both sides of the metasurfaces, for example, the achievable data rate and localization accuracy. In this paper, we focus on STAR-RIS-empowered simultaneous indoor and outdoor three-dimensional (3D) localization, and study the fundamental performance limits via Fisher information analyses and Cram\'er Rao lower bounds (CRLBs). We also devise an efficient localization algorithm based on an off-grid compressive sensing (CS) technique relying on atomic norm minimization (ANM). The impact of the training overhead, the power splitting at the STAR-RIS, the power allocation between the users, the STAR-RIS size, the imperfections of the STAR-RIS-to-BS channel, as well as the role of the multi-path components on the positioning performance are assessed via extensive computer simulations. It is theoretically showcased that high-accuracy, up to centimeter level, 3D localization can be simultaneously achieved for indoor and outdoor users, which is also accomplished via the proposed ANM-based estimation algorithm.   
\end{abstract}

\maketitle

\section{Introduction}
Besides contributing to communications for improved energy efficiency (EE) and spectrum efficiency (SE)~\cite{huang2019reconfigurable,di_renzo_smart_2019,WavePropTCCN, risTUTORIAL2020,RISE6G_COMMAG}, reconfigurable intelligent surfaces (RISs) also play a critical role in radio localization as well as environment mapping, termed as simultaneous localization and mapping (SLAM), in current and upcoming future cellular networks~\cite{wymeersch2019radio, Jiguang2020, Elzanaty2021, Alexandropoulos2022,he2022simultaneous,Tsinghua_RIS_Tutorial}. In these radio localization literature, the RIS behaves in various manners, e.g., a programmable reflector in~\cite{wymeersch2019radio,Jiguang2020,Elzanaty2021}, a cost-efficient receiver in~\cite{Alexandropoulos2022}, and a simultaneous reflector and refractor in~\cite{he2022simultaneous}. In principle, localization performance can be significantly boosted by deploying one or multiple RISs thanks to the subsequent reasons: i) The number of reference nodes, including base stations (BSs), can be further increased with the introduction of RISs; Namely, the RIS can be considered as an additional anchor upon its deployment; Its exact location can be shared with the surrounding BSs. ii) RIS creates a virtual line-of-sight (LoS) link in the millimeter wave (e.g., 5G frequency range 2 (FR2)) network when direct LoS link is temporally unavailable; This happens frequently for millimeter wave communications, known as blockage; iii) Provided that large-sized RISs are exploited, one can obtain high resolution on angular parameters, e.g., angles of departure (AoDs) or angles of arrival (AoAs), associated with RISs for the purpose of user localization; iv) Tremendous RIS beamforming gain leads to enhanced received signal strength, which in turn boosts the localization performance.

Among different types of RIS in~\cite{wymeersch2019radio, Jiguang2020, Elzanaty2021, Alexandropoulos2022,alexandg_2021,he2022simultaneous,Hris_mag}, simultaneously transmitting (a.k.a. refracting) and reflecting RIS (STAR-RIS) stands out as it provides full-dimensional coverage (i.e., $360^{\circ}$ coverage). The application of STAR-RIS for multiple-input multiple-output (MIMO) communications can be referred to~\cite{Yuanwei2021} for a general overview,~\cite{Chenyu2022} for channel estimation, and~\cite{Xinwei2022} for non-orthogonal multiple access (NOMA) transmissions. The STAR-RIS inherently offers two operation functionalities, i.e., reflection and refraction, controlled by two separate series of phase shifters. Such an extraordinary property can also be leveraged for radio localization. For instance, an outdoor BS can simultaneously localize indoor and outdoor mobile stations (MSs) with the aid of the STAR-RIS~\cite{he2022simultaneous}. In this example, the STAR-RIS serves as a tunable reflector for the outdoor MS and meanwhile a tunable refractor for the indoor MS. With the introduced flexibility on power/energy splitting and duplex mode between the two functionalities, the quality-of-service (QoS) requirements in terms of localization accuracy can be met concurrently for both indoor and outdoor MSs.    

To the best of the authors' knowledge, this is the first paper introducing STAR-RIS for simultaneous indoor and outdoor three-dimensional (3D) localization and analyzing the structure's theoretical performance limits~\cite{he2022simultaneous}. However, practical localization algorithms are left undeveloped and several practical issues are left uninvestigated. Thus, in this paper, we continue to focus on the STAR-RIS-enabled simultaneous indoor and outdoor 3D localization system, which comprises one indoor MS and one outdoor MS. The localization of the two users is performed at the BS by considering the received sounding reference signals (SRSs) transmitted over the uplink from the two MSs simultaneously, i.e., in a NOMA manner. We summarize the fundamental performance limits captured by Fisher information analyses and Cram\'er Rao lower bounds (CRLBs), develop effective localization algorithms based on co-channel interference mitigation and off-grid compressive sensing (CS) technique, named atomic norm minimization (ANM), and examine the impact brought by the practical issues, i.e.,  training overhead, power splitting at STAR-RIS and power allocation between the two users, sup-optimal/optimal STAR-RIS design, imperfectness of STAR-RIS-to-BS channel, and multi-path components (MPCs), on the 3D localization performance of the two MSs.  

The rest of the paper is organized as follows. Section~\ref{sec_system_model} introduces the system model, including channel and signal models. Section~\ref{sec_CRLB} summarizes the CRLB analyses on the positioning errors and optimizes the STAR-RIS design. In Section~\ref{sec_loc_algo_design}, we provide the practical localization algorithm based on ANM, followed by numerical study and evaluation of different practical factors in Section~\ref{sec_numerical_results}. Finally, we provide the concluding remarks and point out several future research directions in Section~\ref{sec_conclusions}.

\textit{Notations}: A bold lowercase letter $\ba$ denotes a vector, and a bold capital letter $\bA$ denotes a matrix. $(\cdot)^\mathsf{T}$, $(\cdot)^*$, and $(\cdot)^\mathsf{H}$ denote the matrix or vector transpose, conjugate, and Hermitian transpose, respectively. $(\cdot)^{-1}$ denotes inverse of a matrix, $\mathrm{tr}(\cdot)$ denotes the trace operator, $\mathrm{diag}(\ba)$ denotes a square diagonal matrix with the entries of $\ba$ on its diagonal, 
$\bA \otimes \bB$ and $\bA \diamond \bB$ denote the Kronecker and Khatri-Rao products of $\bA$ and $\bB$, respectively,
$\mathbb{E}[\cdot]$ and $\mathrm{var}(\cdot)$ are the expectation and variance operators, $\mathbf{1}$ is the all-one vector, $\mathbf{0}$ denotes the all-zero vector or matrix, $\bI_{M}$ denotes the $M\times M$ identity matrix, $j = \sqrt{-1}$, $\|\cdot\|_\mathrm{F}$ denotes the Frobenius norm of a matrix, and $\|\cdot\|_2$ denotes the Euclidean norm of a vector. $[\ba]_i$, $[\bA]_{ij}$, and $[\bA]_{i:j, i:j}$ denote the $i$th element of $\ba$, the $(i,j)$th element of $\bA$, and the submatrix of $\bA$ formed by rows $i,i+1, \ldots, j$ and columns $i,i+1, \ldots, j$. Finally, $|\cdot|$ returns the absolute value of a complex number. 

\section{System Model}\label{sec_system_model}
\begin{figure}[t]
	\centering
\includegraphics[width=1\linewidth]{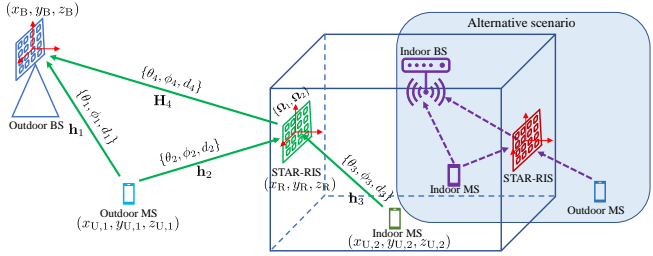}
	\caption{Simultaneous indoor and outdoor 3D localization empowered by the deployment of the STAR-RIS, where an outdoor BS localizes one outdoor MS and one indoor MS concurrently based on the received pilot signals over the uplink.  The two MSs are connected to the outdoor BS via the STAR-RIS with all the links marked with solid arrows. Alternatively, an indoor BS can also localize one indoor MS and one outdoor MS with the help of one STAR-RIS, where all links are marked with dotted arrows.}
		\label{System_Model}
		\vspace{-0.5cm}
\end{figure}
The STAR-RIS-aided 3D localization system, comprising one multi-antenna BS, one multi-element STAR-RIS, one single-antenna indoor MS, and one single-antenna outdoor MS, is depicted in the unshaded area in Fig.~\ref{System_Model}. By post-processing the pilot signals received over the uplink, the outdoor BS is capable of localizing the two MSs simultaneously, termed as simultaneous indoor and outdoor localization. Note that the coverage of the proposed system depends on the BS capabilities and the area of influence of the STAR-RIS \cite{AoI2022}. Alternatively, an indoor BS can also localize one indoor MS and one outdoor MS with the assistance of one STAR-RIS, as depicted in the shaded region in Fig.~\ref{System_Model}. In this paper, we focus on the former scenario and leave the latter for future investigations. 

In the studied 3D localization system, the BS and STAR-RIS are assumed to be equipped with $M$ antennas and $N$ passive scattering elements, respectively. Without loss of generality, we further assume that both the BS and STAR-RIS employ the uniform planar array (UPA) structure parallel to the $x\text{-}z$ plane. They can also be placed parallel to the $y\text{-}z$ plane. In this sense, the corresponding array response vectors discussed in Section~\ref{sec_channel_model} need to be modified accordingly. Recall that STAR-RIS has two operation functionalities, i.e., reflection and refraction, which can be realized simultaneously via two separate series of phase shifters. Therefore, the operation of STAR-RIS can be represented by two independent control matrices, one for controlling reflection and the other for controlling refraction.  

\subsection{Channel Model}\label{sec_channel_model}
The system adopts millimeter wave frequency band for its operation thanks to the availability of substantial spectrum. Thus, we consider the 
Saleh-Valenzuela parametric channel model to construct all the four individual channels, marked by the solid arrows in Fig.~\ref{System_Model}. The direct line-of-sight (LoS) channel between the outdoor MS and the $M$-antenna outdoor BS is denoted as $\bh_1\in\mathbb{C}^{M \times 1}$ and is mathematically expressed as follows:
\begin{equation}\label{h_1}
\bh_1 = \frac{e^{-j 2\pi d_1/\lambda}}{\sqrt{\rho_1}} \boldsymbol{\alpha}_x(\theta_{1},\phi_{1}) \otimes \boldsymbol{\alpha}_z(\phi_{1}),
\end{equation}
 where $d_1$ (in meters) and $\rho_1$ (for the sake of simplicity, we assume that $\rho_1 = d_1^2$) are the distance and path loss between the outdoor MS and the outdoor BS, respectively, $\lambda$ is the wavelength of the carrier frequency, and $\theta_1$ and $\phi_1$ are the azimuth and elevation AoAs associated with $\bh_1$, respectively. In the literature, we can also consider the free-space path loss, modeled as $\rho_1 = d_1^2 f_c^2 / 10^{8.755}$, where $f_c$ (in KHz) is the carrier frequency, defined as $f_c = \frac{c}{\lambda}$ with $c$ being the speed of light. In addition, the standard 3GPP urban micro (UMi) path
loss model can be considered, according to which holds: $\rho_1 = 10^{2.27} d_1^{3.67} f_c^{2.6}$, where $f_c$ needs to be included in GHz~\cite{Akdeniz2014}.

As the BS's antenna array is parallel to $x \text{-} y$ plane, the array response vectors $\boldsymbol{\alpha}_x(\theta_{1},\phi_{1})$ and $\boldsymbol{\alpha}_z(\phi_{1})$ can be written as~\cite{Tsai2018}:
 \begin{align}     \boldsymbol{\alpha}_x(\theta_{1},\phi_{1}) =& \Big[e^{-j \frac{2\pi d_x}{\lambda} (\frac{M_x -1}{2}) \cos(\theta_1) \sin(\phi_1)},\nonumber \\
    & \ldots, e^{j \frac{2\pi d_x}{\lambda} (\frac{M_x -1}{2}) \cos(\theta_1)\sin(\phi_1)} \Big]^{\mathsf{T}},\\
       \boldsymbol{\alpha}_z(\phi_{1}) =& \Big[e^{-j \frac{2\pi d_z}{\lambda} (\frac{M_z -1}{2}) \cos(\phi_1) },\nonumber \\
    & \ldots, e^{j \frac{2\pi d_z}{\lambda} (\frac{M_z -1}{2}) \cos(\phi_1)} \Big]^{\mathsf{T}},
 \end{align}
where $M = M_x M_z$ with $M_x$ and $M_z$ being the number of horizontal and vertical BS antennas, respectively, and $d_x$ and $d_z$ denote their inter-element spacing in the horizontal and vertical axes, which are set as half-wavelength without loss of generality. Similarly, the other two channels linking the two MSs and the STAR-RIS, e.g., $\bh_2 \in \mathbb{C}^{N \times 1}$ and $\bh_3\in \mathbb{C}^{N \times 1}$, can be presented in the same manner, as: 
 \begin{equation}\label{h_2_h_3}
\bh_i = \frac{e^{-j2\pi d_i/\lambda}}{\sqrt{\rho_i}} \boldsymbol{\alpha}_x(\theta_{i},\phi_{i}) \otimes \boldsymbol{\alpha}_z(\phi_{i}),
\end{equation}
for $i = 2$ and $3$, where $N = N_x N_z$ is the number of STAR-RIS elements with $N_x$ and $N_z$ denoting the numbers in the horizontal and vertical axes, respectively. Note that the array response vectors $\boldsymbol{\alpha}_x(\cdot)$ and $\boldsymbol{\alpha}_z(\cdot)$ in~\eqref{h_2_h_3} possess the same format compared to those in~\eqref{h_1} but may differ in dimension if $M_x \neq N_x$ and $M_y \neq N_y$. Finally, $\rho_2$ and $\rho_3$ follow the same assumption as that made for $\rho_1$. 
 
The channel between the STAR-RIS and BS, i.e., $\bH_4 \in \mathbb{C}^{M \times N}$, is expressed as follows:
 \begin{equation}\label{H_4}
\bH_4 = \frac{e^{-j 2\pi d_4/\lambda}}{\sqrt{\rho_4}} \boldsymbol{\alpha}_x(\theta_{4},\phi_{4}) \otimes \boldsymbol{\alpha}_z(\phi_{4})(\boldsymbol{\alpha}_x(\theta_{4},\phi_{4}) \otimes \boldsymbol{\alpha}_z(\phi_{4}))^\mathsf{H},
\end{equation}
provided that the BS and the STAR-RIS are deployed in parallel without any biased orientation in terms of their antenna (element) arrays. As seen from Fig.~\ref{System_Model}, there does not exist a direct LoS path between the outdoor BS and the indoor MS due to the blockage incurred by the wall in between them. The only path connecting them is the refraction route via the STAR-RIS. Unlike the indoor MS, there exist one direct LoS path and one reflection path via the STAR-RIS between the outdoor MS and the outdoor BS.

In the channel model, we consider only LoS paths for all the individual channels; the extension for the multipath scenario are only examined in the numerical study in Section~\ref{sec_numerical_results} by adding random errors. We ignore the possible orientations between the two UPAs (one for BS and the other for STAR-RIS), since they can be known \textit{in priori} and compensated  when implementing practical estimation algorithms for the angular parameters, i.e., azimuth and elevation AoAs.

\subsection{Geometric Relationship}\label{sec_geometric_relationship}
Given a pair of nodes, the geometric relationship is built between their Cartesian coordinates and the latent channel parameters, e.g., $d_1$, $\theta_1$, and $\phi_1$ in~\eqref{h_1}. 
The Cartesian coordinates of the BS and STAR-RIS as well as the outdoor and indoor MSs are $\bp_\text{B} = (x_\text{B},y_\text{B},z_\text{B})^\mathsf{T}$, $\bp_\text{R} = (x_\text{R},y_\text{R},z_\text{R})^\mathsf{T}$, $\bp_{\text{U},1} = (x_{\text{U},1},y_{\text{U},1},z_{\text{U},1})^\mathsf{T}$, and $\bp_{\text{U},2} = (x_{\text{U},2},y_{\text{U},2},z_{\text{U},2})^\mathsf{T}$, respectively. The relationship between the distances and a pair of Cartesian coordinates are listed below:
\begin{align}
    d_1 &= \|\bp_\text{B} - \bp_{\text{U},1} \|_2, \\
     d_i &=  \|\bp_\text{R} - \bp_{\text{U},i-1} \|_2,\;\text{for}\; i = 2,3, \\
       d_4 &=  \|\bp_\text{B} - \bp_{\text{R}} \|_2.
\end{align}
By introducing the three-element vector $\boldsymbol{\xi}_i \triangleq [\cos(\theta_i) \cos(\phi_i),\\ \sin(\theta_i) \cos(\phi_i), \sin(\phi_i) ]^\mathsf{T}$ for $i = 1,2,3,4$, the geometric relationship between the angular parameters and the Cartesian coordinates of the nodes can be expressed as 
\begin{align}\label{Geometry}
    \bp_{\text{R}} &= \bp_{\text{B}} + d_4 \boldsymbol{\xi}_4,  \\  
    \bp_{\text{U},1} & = \bp_{\text{B}} + d_1 \boldsymbol{\xi}_1 = \bp_{\text{R}} + d_2 \boldsymbol{\xi}_2,  \label{p_u_1_1} \\
     \bp_{\text{U},2} & = \bp_{\text{R}} + d_3 \boldsymbol{\xi}_3. \label{p_u_2}
\end{align}

The geometric relationship plays an important role in localization. According to~\eqref{p_u_1_1} and~\eqref{p_u_2}, the BS can calculate the coordinate of the MSs based on the estimates of channel parameters ($d_i, \theta_i, \phi_i$, for $i=1,2,3$), and the pre-known coordinates of the anchors ($\bp_{\text{B}}$ and/or $\bp_{\text{R}}$). 

 \subsection{Signal Model}
It is known that the STAR-RIS has two operation functionalities, i.e., reflection and refraction, which are realized by two separate series of phase shifters. We introduce two phase control matrices, i.e., $\boldsymbol{\Omega}_1 \in \mathbb{C}^{N \times N}$ for controlling refraction and $\boldsymbol{\Omega}_2\in \mathbb{C}^{N \times N}$ for controlling reflection, which are diagonal matrices with each diagonal element satisfying the unit-modulus constraints, e.g., $|[\boldsymbol{\Omega}_1]_{jj}| = |[\boldsymbol{\Omega}_2]_{jj}| = 1$, $\forall j=1,2,\ldots,N$. However, non-ideal reflection and refraction bring attenuation loss, results in reduced modulus, i.e., $|[\boldsymbol{\Omega}_1]_{jj}| <1$ and $|[\boldsymbol{\Omega}_2]_{jj}| < 1$~\cite{Wu2019}.  We consider the 3D localization via the uplink transmission, where the two users send their SRSs towards the BS in a NOMA manner. 
The received signal during the $k$th time slot, for $k =1,2,\ldots,K$, can be mathematically expressed as
 \begin{equation} \label{by_k}
     \by_k = \bh_1 x_{1,k} + \epsilon_2 \bH_4\boldsymbol{\Omega}_{2,k}\bh_2 x_{1,k} + \epsilon_1 \bH_4\boldsymbol{\Omega}_{1,k}\bh_3 x_{2,k} + \bn_k,
 \end{equation}
where $x_{1,k}$ is the SRS from the outdoor MS, $x_{2,k}$ is the SRS from the indoor MS, and coefficients $\epsilon_1$ (for refraction) and $\epsilon_2$ (for reflection) are used to control the power splitting for the two different operational modes of the STAR-RIS, normalized as $\epsilon_1^2 + \epsilon_2^2 = 1$. The received signal at the BS is further corrupted by the white Gaussian noise $\bn_k$, and each element of $\bn_k$ follows complex Gaussian distribution $\mathcal{CN}(0, \sigma^2)$ with zero mean and $\sigma^2$ variance. During the $k$th time slot, the refraction matrix $\boldsymbol{\Omega}_{1,k}$ and the reflection matrix $\boldsymbol{\Omega}_{2,k}$ are considered at the STAR-RIS. In order to ensure good estimates of the channel parameters and the locations of the MSs, $\boldsymbol{\Omega}_{1,k}$ and $\boldsymbol{\Omega}_{2,k}$ vary from one time slot to another, i.e., $\boldsymbol{\Omega}_{1,1} \neq \boldsymbol{\Omega}_{1,2} \neq \ldots \neq  \boldsymbol{\Omega}_{1,K}$ and $\boldsymbol{\Omega}_{2,1} \neq \boldsymbol{\Omega}_{2,2} \neq \ldots \neq  \boldsymbol{\Omega}_{2,K}$. The design of the refractive/reflective beam sweeping will be optimized in Section~\ref{Optimal_Design_of_STAR_RIS} and verified through our numerical results in Section~\ref{sec_numerical_results}.  

The received signal vector $\by_k$ in \eqref{by_k} can be further expressed as 
\begin{align}
    \by_k = &\bh_1 x_{1,k} + \epsilon_2 \bH_4 \mathrm{diag}(\bh_2) \boldsymbol{\omega}_{2,k} x_{1,k} \nonumber\\
    &+ \epsilon_1 \bH_4 \mathrm{diag}(\bh_3)\boldsymbol{\omega}_{1,k} x_{2,k} + \bn_k,
\end{align}
where $\boldsymbol{\Omega}_{1,k} = \mathrm{diag} (\boldsymbol{\omega}_{1,k})$ and  $\boldsymbol{\Omega}_{2,k} = \mathrm{diag} (\boldsymbol{\omega}_{2,k})$, $\forall k$. By stacking all $\by_k$'s column by column, we get the expression:
 \begin{align} \label{bY}
     \bY =& \eta_1 \sqrt{P} \bh_1 \mathbf{1}^\mathsf{T} + \eta_1 \sqrt{P} \epsilon_2 \bH_4 \mathrm{diag}(\bh_2) \bar{\boldsymbol{\Omega}}_2 \nonumber \\
     &+ \eta_2 \sqrt{P} \epsilon_1 \bH_4 \mathrm{diag}(\bh_3)\bar{\boldsymbol{\Omega}}_1 + \bN,
 \end{align}
where $\mathbf{1}$ denotes the $K$-element all-one vector, $\bY = [\by_1, \ldots, \by_K]$, $\bN = [\bn_1, \ldots, \bn_K]$, $\bar{\boldsymbol{\Omega}}_1 = [\boldsymbol{\omega}_{1,1}, \ldots, \boldsymbol{\omega}_{1,K}]$, and $\bar{\boldsymbol{\Omega}}_2 = [\boldsymbol{\omega}_{2,1}, \ldots, \\ \boldsymbol{\omega}_{2,K}]$ with $|[\bar{\boldsymbol{\Omega}}_1 ]_{mn}| = |[\bar{\boldsymbol{\Omega}}_2 ]_{mn}| =1, \forall m, n$. Without loss of generality, we assume that the sum transmit power constraint is applied for each time slot, i.e., $|x_{1,k}|^2 + |x_{2,k}|^2 = P$, $\forall k$, and introduce coefficients $\eta_1$ and $\eta_2$ for characterizing the power allocation between the two MSs, satisfying $|x_{1,k}|^2 = \eta_1^2 P$,  $|x_{2,k}|^2 = \eta_2^2 P$, and $\eta_1^2 + \eta_2^2 =1$. Based on the received signals across $K$ time slots, the BS estimates the Cartesian coordinates of both the indoor and outdoor users, enabling 3D localization.

Applying vectorization to $\bY$ in~\eqref{bY}, we get the following expression:
 \begin{align}\label{vec_Y}
     \by =& \eta_1\sqrt{P} (\mathbf{1} \otimes \bI_{M}) \bh_1+ \eta_1 \sqrt{P} \epsilon_2 (\bar{\boldsymbol{\Omega}}_2^\mathsf{T} \otimes \bI_M)(\bI_N \diamond \bH_4) \bh_2\nonumber\\
     &+ \eta_2 \sqrt{P} \epsilon_1 (\bar{\boldsymbol{\Omega}}_1^\mathsf{T} \otimes \bI_M)(\bI_N \diamond \bH_4) \bh_3 + \bn,
 \end{align}
 where $\by = \mathrm{vec}(\bY)$ and $\bn = \mathrm{vec}(\bN) \sim \mathcal{CN}(\boldsymbol{0}, \sigma^2\bI_{KM})$. The expression in~\eqref{vec_Y} can be re-written as: 
 \begin{equation} \label{vec_Y1}
     \by =\sqrt{P} \eta_1\bA_1 \bh_1 +  \sqrt{P}\eta_1 \epsilon_2 \bA_2 \bh_2 + \sqrt{P}\eta_2\epsilon_1\bA_3 \bh_3 + \bn, 
 \end{equation}
 by introducing the following three new notations:
\begin{align} \label{A_1}
    \bA_1 &=  (\mathbf{1} \otimes \bI_M) \in \mathbb{C}^{KM \times M}, \\
    \bA_2 &=   (\bar{\boldsymbol{\Omega}}_2^\mathsf{T} \otimes \bI_M)(\bI_N \diamond \bH_4)\in \mathbb{C}^{KM \times N}, \\
    \bA_3 & =    (\bar{\boldsymbol{\Omega}}_1^\mathsf{T} \otimes \bI_M)(\bI_N \diamond \bH_4)\in \mathbb{C}^{KM \times N}. \label{A_3}
\end{align}
As we can see from~\eqref{A_1} to~\eqref{A_3}, $\bA_1$ is independent of the STAR-RIS design, while $\bA_2$ and $\bA_3$ are functions of $\bar{\boldsymbol{\Omega}}_2$ and $\bar{\boldsymbol{\Omega}}_1$, respectively. 

Upon the STAR-RIS deployment, we assume that the BS knows the exact/precise location of the STAR-RIS. Thus, we assume that the BS has exact information on $\bH_4$ in terms of the parameters $\theta_4$, $\phi_4$, and $d_4$. Thus, $\bA_1$, $\bA_2$, and $\bA_3$ in~\eqref{vec_Y1} are known measurement matrices to the BS (the BS also knows the refractive/reflective phase configurations due to its interaction with the STAR-RIS controller) in the theoretical performance limit analyses in Section~\ref{sec_CRLB} and localization algorithm development in Section~\ref{sec_loc_algo_design}. However, this assumption will be relaxed and its effect will be examined in Section~\ref{sec_numerical_results} since perfect information on $\bH_4$ is usually infeasible in practice.

 \section{Cram\'er Rao Lower Bound Analyses}\label{sec_CRLB}
In this section, we summarize the CRLBs on the estimation of the intermediate channel parameters from~\cite{he2022simultaneous}, followed by the 3D Cartesian coordinates' estimation. This two-step approach can be commonly seen in the literature~\cite{Jiguang2020,Elzanaty2021,Shahmansoori2017}. We also present the refraction/reflection optimization of the STAR-RIS for the 3D localization objective, where the case $K \geq 2 N +1$ is considered and its optimal solution is found. \footnote{In general, a reasonable training overhead is required in order to achieve satisfactory localization performance for both users. Thus, in this work, we only focus on the scenario with $K$ slightly larger than $2 N +1$, which fits well with the aforementioned statement. As said, we can find the optimal STAR-RIS design for such a case, detailed in Section~\ref{Optimal_Design_of_STAR_RIS}.} 
 
\subsection{Estimation of Channel Parameters}
The unknown channel parameters to be estimated are those included in $\bh_1$, $\bh_2$, and $\bh_3$, i.e., the nine-tuple $\boldsymbol{\nu} \triangleq [\theta_1, \phi_1, d_1,\theta_2, \phi_2, d_2,\theta_3, \\ \phi_3, d_3]^\mathsf{T}$. Since the additive noise follows complex Gaussian distributed, by introducing $\boldsymbol{\mu}(\boldsymbol{\nu}) \triangleq  \eta_1\bA_1 \bh_1 +  \eta_1 \epsilon_2 \bA_2 \bh_2 + \eta_2\epsilon_1\bA_3 \bh_3$ from~\eqref{vec_Y1}, the Fisher information matrix for $\boldsymbol{\nu}$ is obtained as:
\begin{equation}\label{Fisher_parameter}
[\bJ(\boldsymbol{\nu}) ]_{i,j} =  \frac{P}{\sigma^2}\Re \Big\{\frac{\partial \boldsymbol{\mu}^\mathsf{H}} {\partial \nu_i} \frac{ \partial \boldsymbol{\mu}}{ \partial \nu_j} \Big\}.
\end{equation}
The information on the partial derivatives in~\eqref{Fisher_parameter} related to parameters in $\bh_1$, $\bh_2$, and $\bh_3$ can be referred to~\cite{he2022simultaneous} for more details. 

For any unbiased estimator (denoted by $\hat{\boldsymbol{\nu}}(\by)$) for the channel parameters, we can calculate the CRLB on the error covariance matrix as follows:
\begin{equation}\label{CRLB_NU}
    \mathbb{E}\{(\boldsymbol{\nu}- \hat{\boldsymbol{\nu}}(\by))(\boldsymbol{\nu}- \hat{\boldsymbol{\nu}}(\by))^\mathsf{H} \} \succeq \bJ^{-1}(\boldsymbol{\nu}),
\end{equation}
where the notation $\bA \succeq \bB$ for square matrices $\bA$ and $\bB$ means $\ba^\mathsf{H} \bA \ba \geq \ba^\mathsf{H} \bB \ba$ for any valid vector $\ba$. The expression~\eqref{CRLB_NU} indicates that the estimation error variance for each individual channel parameter in $\boldsymbol{\nu}$ is lower bounded by the corresponding diagonal element in $\bJ^{-1}(\boldsymbol{\nu})$, which indicates the best performance any unbiased estimator can reach in theory.  

\subsection{Estimation of 3D Cartesian Coordinates}
Our ultimate goal is to estimate the 3D Cartesian coordinates of the indoor and outdoor MSs. Therefore, after estimating the channel parameters, we need to map them to 3D Cartesian coordinates, e.g., $\boldsymbol{\kappa} = [x_{\text{U},1}, y_{\text{U},1}, z_{\text{U},1},x_{\text{U},2}, y_{\text{U},2}, z_{\text{U},2}]^\mathsf{T}$, based on the geometrical relationship among the BS, the STAR-RIS, and the two MSs, discussed in Section~\ref{sec_geometric_relationship}. For the CRLB evaluation of $\boldsymbol{\kappa}$, we resort to the Jacobian matrix $\bT$, which links the connection between the channel parameters $ \boldsymbol{\nu}$ and the 3D Cartesian coordinates of the two MSs $\boldsymbol{\kappa}$. Each $(i,j)$th element of $\bT$ is expressed as:
\begin{equation} \label{T_matrix}
    [\bT]_{ij} = \frac{\partial [\boldsymbol{\nu}]_j}{\partial [\boldsymbol{\kappa}]_i}. 
\end{equation}
Again, we omit the details on the calculation of each individual derivative in~\eqref{T_matrix}, which can be found in~\cite{he2022simultaneous}.

In addition, it can be easily seen that only the channel parameters $\{ \theta_1, \phi_1, d_1,\theta_2, \phi_2, d_2\}$ are related to the coordinates $(x_{\text{U},1}, y_{\text{U},1}, z_{\text{U},1})$ of the outdoor MS, and only the parameters $\{\theta_3, \phi_3, d_3\}$ are related to the coordinates $(x_{\text{U},2}, y_{\text{U},2}, z_{\text{U},2})$ of the indoor MS, as concluded from~\eqref{p_u_1_1} and \eqref{p_u_2}. Therefore, the Jacobian matrix $\bT$ has the following form:
\begin{equation}
\bT = 
\begin{bmatrix}
\bT_{1} & \mathbf{0} \\
\mathbf{0} & \bT_{2} 
\end{bmatrix},
\end{equation} 
where the submatrix $\bT_{1} \in \mathbb{R}^{3\times 6}$ consists of the partial derivatives related to the outdoor MS, and the submatrix $\bT_{2}\in \mathbb{R}^{3\times 3}$ consists of the partial derivatives related to the indoor MS.

The Fisher information of $\boldsymbol{\kappa}$ can be then expressed as~\cite{Elzanaty2021} 
\begin{equation}\label{J_kappa}
    \bJ(\boldsymbol{\kappa}) = \bT  \bJ(\boldsymbol{\nu}) \bT^\mathsf{T}. 
\end{equation}
Similar to~\eqref{CRLB_NU}, we have the inequality for the CLRB:
\begin{equation}\label{CRLB_kappa}
    \mathbb{E}\{(\boldsymbol{\kappa}- \hat{\boldsymbol{\kappa}}(\by))(\boldsymbol{\kappa}- \hat{\boldsymbol{\kappa}}(\by))^\mathsf{T} \} \succeq \bJ^{-1}(\boldsymbol{\kappa}).
\end{equation}
The performance lower bounds on the root mean square error (RMSE) of the position estimation of the outdoor and indoor MSs are:
\begin{align}
 \text{RMSE}_{\text{U},1} = \sqrt{\mathrm{var}(\hat{\bx}_{\text{U},1})} &\geq \sqrt{\mathrm{tr}\{[\bJ^{-1}(\boldsymbol{\kappa})]_{1:3,1:3}\}}, \\
 \text{RMSE}_{\text{U},2} = \sqrt{\mathrm{var}(\hat{\bx}_{\text{U},2})} &\geq \sqrt{\mathrm{tr}\{[\bJ^{-1}(\boldsymbol{\kappa})]_{4:6,4:6}\}}
\end{align}
where $\hat{\bx}_{\text{U},1}$ and $\hat{\bx}_{\text{U},2}$ are the unbiased estimates of $\bx_{\text{U},1}$ and $\bx_{\text{U},2}$, respectively. 

\subsection{Localization-Optimal Design for the STAR-RIS}\label{Optimal_Design_of_STAR_RIS}
In this subsection, we consider the optimization of STAR-RIS during the pilot transmission, aiming at maximizing the overall 3D localization performance of the two MSs. The scenario with $K \geq 2N +1$ is considered, where we aim at optimizing the STAR-RIS based on the inverse matrix of the Fisher information matrix. By introducing $\bG_1 \triangleq [\frac{\partial \boldsymbol{\mu}} {\partial \theta_1}, \frac{\partial \boldsymbol{\mu}} {\partial \phi_1}, \frac{\partial \boldsymbol{\mu}} {\partial d_1}, \frac{\partial \boldsymbol{\mu}} {\partial \theta_2},  \frac{\partial \boldsymbol{\mu}} {\partial \phi_2}, \frac{\partial \boldsymbol{\mu}} {\partial d_2} ]$, $\hat{\bG}_1 \triangleq \bG_1 \bT_{1}^\mathsf{H}$, $\bG_2 \triangleq [\frac{\partial \boldsymbol{\mu}} {\partial \theta_3}, \frac{\partial \boldsymbol{\mu}} {\partial \phi_3}, \frac{\partial \boldsymbol{\mu}} {\partial d_3}]$, and $\hat{\bG}_2 \triangleq \bG_2 \bT_{2}^\mathsf{H}$, the expression of $\bJ^{-1}(\boldsymbol{\kappa})$ in~\eqref{CRLB_NU} can be expressed as~\cite{scharf1993geometry, Pakrooh2015}

\begin{equation}\label{bJ_inv}
      \bJ^{-1}\!(\boldsymbol{\kappa}) \!=\! \frac{\sigma^2}{P}
      \begin{bmatrix}
     (\hat{\bG}_1^\mathsf{H} (\bI \!-\! \bP_{\hat{\bG}_2})\hat{\bG}_1)^{-1} & * \\
      * &  (\hat{\bG}_2^\mathsf{H} (\bI \!-\! \bP_{\hat{\bG}_1})\hat{\bG}_2 )^{-1}
      \end{bmatrix}\!\!,
\end{equation}
where $\bP_{\hat{\bG}_i} = \hat{\bG}_i( \hat{\bG}_i^\mathsf{H} \hat{\bG}_i)^{-1}\hat{\bG}_i^\mathsf{H} $ is the orthogonal projection onto the column space of $\hat{\bG}_i$ for $i = 1$ and $2$.

In order to simplify the diagonal terms in~\eqref{bJ_inv}, we rewrite $\bG_1$ and $\bG_2$ as $\bG_1 = [\eta_1\bA_1\bQ_1, \; \eta_1 \epsilon_2 \bA_2  \bQ_2]$ and $\bG_2 = \eta_2 \epsilon_1 \bA_3\bQ_3$, where $\bQ_1 \in\mathbb{C}^{M\times 3}$, $\bQ_2\in\mathbb{C}^{N\times 3}$, and $\bQ_3\in\mathbb{C}^{N\times 3}$ contain the remaining terms of $\{\frac{\partial \boldsymbol{\mu}} {\partial \theta_1}, \frac{\partial \boldsymbol{\mu}} {\partial \phi_1}, \frac{\partial \boldsymbol{\mu}} {\partial d_1}\}, \{\frac{\partial \boldsymbol{\mu}} {\partial \theta_2},  \frac{\partial \boldsymbol{\mu}} {\partial \phi_2}, \frac{\partial \boldsymbol{\mu}} {\partial d_2}\}, \{\frac{\partial \boldsymbol{\mu}} {\partial \theta_3},  \frac{\partial \boldsymbol{\mu}} {\partial \phi_3},\\ \frac{\partial \boldsymbol{\mu}} {\partial d_3}\}$ (excluding $\eta_1\bA_1$, $\eta_1 \epsilon_2\bA_2$, and $\eta_2 \epsilon_1 \bA_3$), respectively. Our goal is to optimize $\bar{\boldsymbol{\Omega}}_1$ in $\bA_3$ and $\bar{\boldsymbol{\Omega}}_2$ in $\bA_2$ in order to obtain the best theoretical localization performance.  By further dividing $\bT_{1}$ into two submatrices, i.e., as $\bT_{1} = [\tilde{\bT}_{1}\in \mathbb{R}^{3\times 3}, \bar{\bT}_{1}\in \mathbb{R}^{3\times 3}]$, we can derive the expressions $\hat{\bG}_1 = \eta_1\bA_1\bQ_1\tilde{\bT}_{1}^\mathsf{H} +  \eta_1 \epsilon_2 \bA_2  \bQ_2\bar{\bT}_{1}^\mathsf{H}$ and $\hat{\bG}_2 = \eta_2 \epsilon_1 \bA_3\bQ_3\bT_{2}^\mathsf{H}$. For the sake of tractability for the STAR-RIS optimization (i.e., $\bar{\boldsymbol{\Omega}}_1$ and $\bar{\boldsymbol{\Omega}}_2$), we maximize the principal angle (within $[0,\pi/2]$) between the subspaces of $\hat{\bG}_1$ and $\hat{\bG}_2$ by following~\cite{scharf1993geometry}, which is equivalent to minimizing $\| \hat{\bG}_2 ^\mathrm{H} \hat{\bG}_1\|_\mathrm{F}$, i.e., 
\begin{equation}\label{Frobenious_norm}
\big\| \eta_1 \eta_2 \epsilon_1 \bT_2 \bQ_3^\mathrm{H} \bA_3^\mathrm{H}  \bA_1\bQ_1\tilde{\bT}_{1}^\mathsf{H} + \eta_1 \epsilon_2 \eta_2 \epsilon_1 \bT_2 \bQ_3^\mathrm{H} \bA_3^\mathrm{H}\bA_2  \bQ_2\bar{\bT}_{1}^\mathsf{H}\big\|_\mathrm{F}.  
\end{equation}
We focus on $\bA_3^\mathrm{H}  \bA_1$ in the first term and $\bA_3^\mathrm{H}  \bA_2$ in the second term of~\eqref{Frobenious_norm}, which are detailed as  
\begin{align}\label{A_3_times_A_1}
\bA_3^\mathrm{H}  \bA_1 &= (\bI_N \diamond \bH_4^\mathrm{H} ) (\bar{\boldsymbol{\Omega}}_1^*\mathbf{1}  \otimes \bI_M ), \\
\bA_3^\mathrm{H}  \bA_2 &= (\bI_N \diamond \bH_4^\mathrm{H} ) (\bar{\boldsymbol{\Omega}}_1^*\bar{\boldsymbol{\Omega}}_2^\mathrm{T}  \otimes \bI_M )(\bI_N \diamond \bH_4 ),\label{A_3_times_A_2}
\end{align}
by using the following properties: $(\bB \bC)^\mathrm{H} = \bC^\mathrm{H} \bB^\mathrm{H}$, $(\bB \times \bC)(\bD \times \bE) = (\bB \bD \otimes \bC \bE)$, $ (\bB \otimes \bC)^\mathrm{H} = \bB^\mathrm{H} \otimes \bC^\mathrm{H}$, and $(\bB \diamond \bC)^\mathrm{H} = \bB^\mathrm{H} \diamond \bC^\mathrm{H}$ for matrices $\bB$, $\bC$, $\bD$, and $\bE$ meeting the dimension requirements. 

When training overhead $K \geq 2N +1$, we can easily find the optimal solution for $ \bar{\boldsymbol{\Omega}}_1$ and $\bar{\boldsymbol{\Omega}}_2$ that minimize the Frobenius norm in~\eqref{Frobenious_norm}, e.g., $ \bar{\boldsymbol{\Omega}}_1 = [\bW]_{2:N+1,:}$ and $ \bar{\boldsymbol{\Omega}}_2 = [\bW]_{N+2:2N+1, :}$ with $ \bW$ being the transpose of a $K \times K$ discrete Fourier transform (DFT) matrix or Hadamard matrix (only for certain integer $K$ values). By following this, we have $\bar{\boldsymbol{\Omega}}_1^*\mathbf{1}  = \mathbf{0}$ in~\eqref{A_3_times_A_1} and $\bar{\boldsymbol{\Omega}}_1^*\bar{\boldsymbol{\Omega}}_2^\mathrm{T} = \mathbf{0}$ in~\eqref{A_3_times_A_2}, which in turn yields $\| \hat{\bG}_2 ^\mathrm{H} \hat{\bG}_1\|_\mathrm{F} = 0$ (i.e., we get the largest principal angle between the subspaces of $\hat{\bG}_1$ and $\hat{\bG}_2$, which is  $\pi/2$~\cite{golub2013matrix}). 


\section{Localization Algorithm Development}\label{sec_loc_algo_design}
By aligning with the theoretical analyses studied in Section~\ref{sec_CRLB}, we develop a two-step localization algorithm to estimate the location of the two MSs. In the first step, based on~\eqref{vec_Y1} one estimates the channel parameters in $\bh_1$, $\bh_2$, and $\bh_3$. In the second step, by following the geometric relationship in Section~\ref{sec_geometric_relationship}, we map the estimates of channel parameters to the coordinates of the MSs. Special attention needs to be paid to the first step for the co-channel interference mitigation/cancellation, i.e., null the other two when estimating one channel. Also, for the channel parameter estimation, we resort to off-grid CS ANM for recovering the sparsity-one channel vectors and root multiple signal classification (root-MUSIC) for extracting the corresponding angular channel parameters, followed by distance estimates using the least squares (LS) principle~\cite{he20223d}. For the location mapping of the outdoor MS, we adopt a weighted sum criterion based on the two distance estimates associated with $\bh_1$ and $\bh_2$. 

\subsection{Co-Channel Interference Cancellation}
As an essential process for channel parameter estimation, we need to null the interference for each MS in~\eqref{vec_Y1}. By multiplying the interference nulling matrix $\bU_i$ to $\by$, for $i = 1,2,3$, we get the following 
\begin{equation}\label{sig_after_inter_cancel}
\bU_i \by = \gamma_i \bU_i\bA_i \bh_i + \bU_i\bn, 
\end{equation}
where $\gamma_1 = \sqrt{P}  \eta_1$, $\gamma_2 =\sqrt{P}\eta_1 \epsilon_2$, and $\gamma_3 = \sqrt{P}\eta_2\epsilon_1$. We rely on~\eqref{sig_after_inter_cancel} in the next subsection for channel parameter estimation. The co-channel interference can be completely cancelled in~\eqref{sig_after_inter_cancel} in certain situations depending on the relationship among the vector spaces spanned by the column vectors of $\bA_1$, $\bA_2$, and $\bA_3$, respectively~\cite{Sung2010}. Otherwise, we merge the residual interference to the additive noise term $\bU_i\bn$. One way to design $\bU_1$ is to perform singular value decomposition (SVD) on the combined matrix $[\bA_2, \bA_3]$ and take the left singular vectors corresponding to zero singular values~\cite{Sung2010}. A similar approach can be applied for constructing $\bU_2$ and $\bU_3$. 

\subsection{Channel Parameter Estimation}
All the channel vectors in~\eqref{vec_Y1} are in the form of Kronecker product of two array response vectors. It can be seen as sparsity-one signal (i.e., a linear combination of one atom having the same structure as the channel component in the form of Kronecker product of two array response vectors), which is an extreme case under the framework of CS. Therefore, we follow the off-grid CS techniques, i.e., ANM, for recovering the sparsity-one channel vector. This technique has been used in the literature~\cite{he2020anm, he20223d} for channel estimation and localization purposes. We first introduce the atomic set~\cite{ Yang2016,Tang2013, he2020anm}, as \begin{equation}\label{atomic_set}
\mathcal{A} \triangleq \{\boldsymbol{\alpha}_x(x_1, x_2) \otimes \boldsymbol{\alpha}_z(x_2), x_1 \in [0, \pi], x_2 \in [-\pi/2, \pi/2]  \},  
\end{equation}
where each atom possesses the same structure with the linear term in $\bh_i$, for $i = 1,2,3$. For any vector $\bh_i$ of the form $\bh_i = \sum_{l} \eta_l \boldsymbol{\alpha}_x(x_{1,l}, x_{2,l}) \otimes \boldsymbol{\alpha}_z(x_{2,l})$ with each $\eta_l>0$ being a coefficient, $x_{1,l} \in [0, \pi]$, and $x_{2,l} \in [-\pi/2, \pi/2]$, its atomic norm with respect to the atomic set $\mathcal{A}$ is written as follows:
\begin{align}
\|\bh_i\|_{\mathcal{A}} = &\mathrm{inf}_{\mathcal{B}}\Big\{\frac{1}{2T_i} \mathrm{Tr}(\mathrm{Toep}(\mathcal{U}_2)) + \frac{t}{2}\Big\}, \nonumber\\
&\text{s.t.} \;\begin{bmatrix} \mathrm{Toep}(\mathcal{U}_2)  & \bh_i\\
\bh_i^{\mathsf{H}}& t_i
\end{bmatrix} \succeq \mathbf{0},
\end{align}
where set $\mathcal{B}\triangleq\{\mathcal{U}_2 \in \mathbb{C}^{T_i \times T_i}, t_i\in \mathbb{R}\}$ with $\mathcal{U}_2$ being a $2$-way tensor and $\mathrm{Toep}(\mathcal{U}_2)$ is a $2$-level block Toeplitz matrix, which results from the Vandermonde decomposition lemma for positive semidefinite Toeplitz matrices~\cite{Tang2013}. The value for $T_i$ depends on the dimension of $\bh_i$, i.e., $T_i = M$ for $i =1$, and $T_i = N$, for $i =2,3$.

The ANM based channel estimation can be formulated as a regularized optimization problem:
\begin{align}\label{ANM_h_i}
\hat{\bh}_i = \arg\min_{\bh_i \in \mathbb{C}^{T_i},\; \mathcal{B}} &\mu_i \|\bh_i\|_{\mathcal{A}} + \frac{1}{2}\|\bU_i \by - \gamma_i \bU_i\bA_i \bh_i\|_2^2 \nonumber\\
&\text{s.t.} \;\begin{bmatrix} \mathrm{Toep}(\mathcal{U}_2)  & \bh_i\\
\bh_i^{\mathsf{H}}& t_i
\end{bmatrix} \succeq \mathbf{0},
\end{align}
where $\mu_i \propto \sigma \sqrt{ T_{i} \log(T_{i})}$ is the regularization term of the atomic norm penalty, and $\hat{\bh}_i$ is the estimate of $\bh_i$. This problem can be efficiently solved using the Matlab CVX toolbox. 

Based on the $\hat{\bh}$,  the elevation and azimuth AoAs can be extracted by following root-MUSIC algorithm~\cite{he20223d} and the distance $d_i$ can be estimated by following LS principle, as  
\begin{equation}\label{estimate_d_i}
\hat{d}_i = \sqrt{T_i/\bh_i^\mathsf{H} \bh_i },
\end{equation}
where $\hat{d}_i$ is the estimate of $d_i$. 

\subsection{Location Mapping}

Based on the estimate of $\bh_i$ in~\eqref{ANM_h_i}, we can further resort to root-MUSIC for extracting the angular parameters, denoted by $\hat{\theta}_i$ and $\hat{\phi}_i$~\cite{he20223d}. Together with the estimate of $d_i$ in~\eqref{estimate_d_i}, the location of the MSs can be calculated by following the geometric relationship in Section~\ref{sec_geometric_relationship}. 

Specifically, for the location estimate of the outdoor MS, since both $\bh_1$ and $\bh_2$ contribute to it, we apply weighted sum principle, as 
\begin{equation}
\bp_{\text{U},1}  = w_1 (\bp_{\text{B}} + \hat{d}_1 \hat{\boldsymbol{\xi}}_1) + (1-w_1)(\bp_{\text{R}} + \hat{d}_2 \hat{\boldsymbol{\xi}}_2),
\end{equation}
where the weight $w_1 = \frac{\hat{d}_2^2}{\hat{d}_1^2+\hat{d}_2^2}$ is set in a heuristic way by following that the weight is reversely proportional to the path loss. 

\section{Numerical Results}\label{sec_numerical_results}
In this section's numerical investigation, we set the system parameters as follows: $\bp_\text{B} =(0, 0, 8)^\mathsf{T}$, $\bp_\text{R} =(2, 2, 5)^\mathsf{T}$, $\bp_{\text{U},1} =(5, 1, 2)^\mathsf{T}$, and $\bp_{\text{U},2} =(1, 5, 2)^\mathsf{T}$. The numbers of BS antennas, STAR-RIS elements, and SRSs from each MS are set as $M = 16$, $N = 36$, and $K = 100, 130$. The signal-to-noise ratio (SNR) is defined as $P/\sigma^2$. 
The parameter setup is summarized in Table~\ref{tab1:parameter}.

\begin{table}[t]
    \centering
  \caption{Parameter Setup.}
    \label{tab1:parameter}
    \begin{tabular}{cc|cc}
        \hline
        Parameter  & Value & Parameter  & Value \\
      \hline
        $M$ & $16$ &   $N$ & $36$ \\
        $K$&$100,130$& $\bp_\text{B}$&$(0, 0, 8)^\mathsf{T}$\\
        $\bp_\text{R}$  &  $(2, 2, 5)^\mathsf{T}$  &
        $\bp_{\text{U},1}$ & $(5, 1, 2)^\mathsf{T}$  \\ 
        $\bp_{\text{U},2}$  &  $(1, 5, 2)^\mathsf{T}$ \\
      \hline
    \end{tabular}
\end{table}

\subsection{Effect of Training Overhead}
\begin{figure}[t]
	\centering
\includegraphics[width=0.95\linewidth]{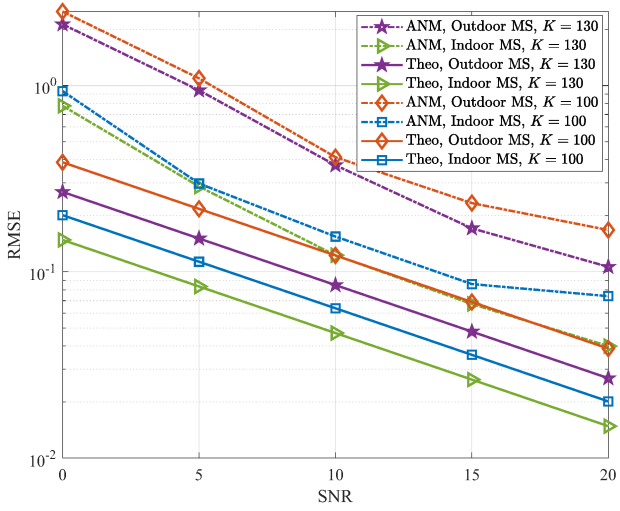}
	\caption{The effect of training overhead on 3D localization with $\epsilon_1 = \sqrt{0.9}$ and $\eta_1 = \sqrt{0.5}$. }
		\label{Effect_of_training_overhead}
\end{figure}

As shown in Section~\ref{Optimal_Design_of_STAR_RIS}, we find the optimal design of the STAR-RIS for the training overhead $K \geq 2N +1$. We pick up two $K$ values meeting this requirement and compare their impact on the localization performance. The simulation results, including both theoretical and practical, are shown in Fig.~\ref{Effect_of_training_overhead} for the training overhead $K = 100, 130$, where $\epsilon_1 = \sqrt{0.9}, \eta_1 = \sqrt{0.5}$. In the legend, ``ANM'' denotes the proposed ANM based 3D localization scheme while ``Theo'' stands for the theoretical performance limit analyses. From the theoretical ones characterized by CRLBs, we know that higher training overhead can bring better localization performance, up to centimeter level for both MSs. Moreover, the indoor MS can achieve better performance than the outdoor MS in such a unbalanced setup on $\epsilon_1$ and $ \eta_1$ since the STAR-RIS power splitting coefficient is large (Namely, more percent of energy is refracted towards the BS via the STAR-RIS). The practical results from ANM are consistent with the theoretical studies. The performance gain brought by increasing the training overhead from $100$ to $130$ is not so obvious from the practical results, especially in the low SNR regime. However, a constant gain (roughly $4$ dB) is observed from the theoretical CRLB results across all the SNR values for both users.

\subsection{Effect of Power Splitting and Allocation}
\begin{figure}[t]
	\centering
\includegraphics[width=0.95\linewidth]{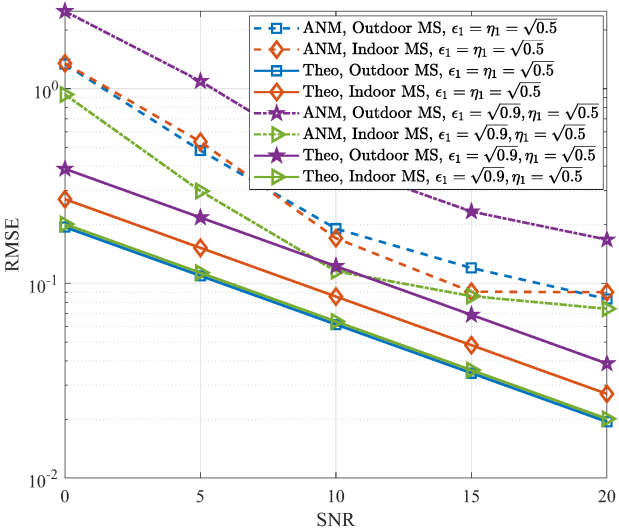}
	\caption{3D localization performance with different pairs of $\epsilon_1$ and $\eta_1$, where the training overhead is set as $K = 100$. }
		\label{Localization_results_low_training_overhead}
\end{figure}

The performance of 3D localization is not only affected by the training overhead but also the two parameters, i.e., $\epsilon_1$ for controlling power splitting at STAR-RIS and $\eta_1$ for controlling power allocation between the two users. For the training overhead $K = 100$, the simulation results with different setups on $\epsilon_1$ and $\eta_1$ are shown in Fig.~\ref{Localization_results_low_training_overhead}. As we can see that when a balanced setup is adopted, i.e., $\epsilon_1 = \eta_1 =\sqrt{0.5}$, the performance gap between the two MSs' position estimation is small. However, in the other setup, i.e., $\epsilon_1 = \sqrt{0.9}, \eta_1 = \sqrt{0.5}$,  the gap is obvious. By carefully choosing the values for the two parameters, we can simultaneously achieve the QoS requirements for both users.

 \subsection{Effect of STAR-RIS Design}
\begin{figure}[t]
	\centering
\includegraphics[width=0.95\linewidth]{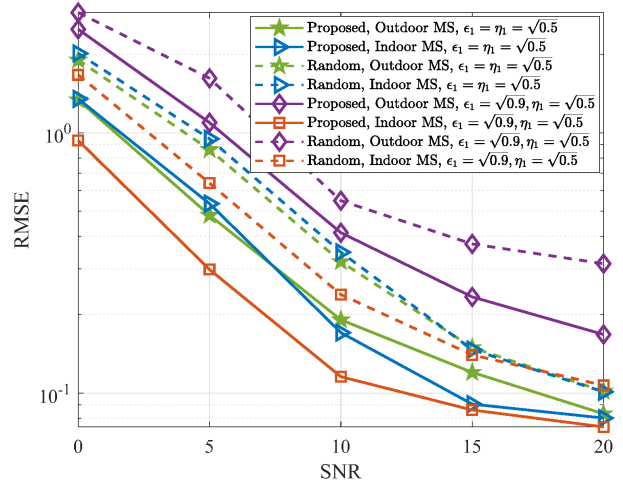}
	\caption{The effect of STAR-RIS design on 3D localization with training overhead $K = 100$ and different pairs of $\epsilon_1$ and $\eta_1$. }
		\label{practical_localization_results_low_training_overhead_STAR_RIS_design}
\end{figure}

In this subsection, we evaluate the effect of the STAR-RIS design (under the condition of training overhead $K =100$) by considering two different cases: i) $\bar{\boldsymbol{\Omega}}_1$ and $\bar{\boldsymbol{\Omega}}_2$ are designed according Section~\ref{Optimal_Design_of_STAR_RIS}; ii) the phases of  $\bar{\boldsymbol{\Omega}}_1$ and $\bar{\boldsymbol{\Omega}}_2$ are randomly generated. The simulation results are shown in Fig.~\ref{practical_localization_results_low_training_overhead_STAR_RIS_design}. As expected, the first case outperforms the second one, since according to Section~\ref{Optimal_Design_of_STAR_RIS} it is optimal. The performance gain between the two cases is obvious with different setups for $\epsilon_1$ and $\eta_1$. For instance, when $\epsilon_1 = \sqrt{0.9}$ and $ \eta_1 = \sqrt{0.5}$, more than $5$ dB gain in terms of SNR can be obtained for the indoor MS by following the optimal STAR-RIS design compared to the random phase design. 
 
 \subsection{Effect of Imperfect STAR-RIS-to-BS Channel}\label{Imperfect_SATR_RIS_to_BS_Channel}
\begin{figure}[t]
	\centering
\includegraphics[width=0.97\linewidth]{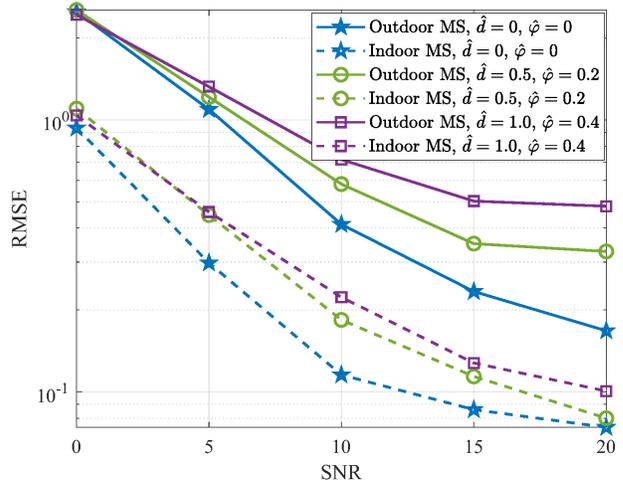}
	\caption{The effect of imperfect STAR-RIS-to-BS channel on 3D localization with training overhead $K = 100$ and different pairs of $\hat{d}$ and $\hat{\varphi}$. }
		\label{imperfect_star_ris_to_BS}
\end{figure}

In this section, we examine the effect of imperfectness of the STAR-RIS-to-BS channel on the 3D localization performance. Different from the previous subsections, here, we assume that the STAR-RIS-to-BS channel matrix is available but in the form of imperfectness during the localization process. We introduce individual random variation to each of the channel parameters in the STAR-RIS-to-BS channel, e.g., $d_4$, $\theta_4$, $\phi_4$. Such variations can be introduced by practical algorithms for estimating them. We further assume that the variations follow uniform distribution, i.e., $\Delta d_4 \sim\mathcal{U}[-\hat{d}, \hat{d}]$, $\Delta \theta_4, \Delta \phi_4  \sim\mathcal{U}[-\hat{\varphi},\hat{\varphi}]$. The values of $\hat{d}$ and $\hat{\varphi}$ jointly determine the level of imperfectness of the STAR-RIS-to-BS channel. In this experiment, we evaluate two cases: i) $\hat{d} = 0.5$ in meter and $\hat{\varphi} = 0.2$ in radian, ii)  $\hat{d} = 1$ in meter and $\hat{\varphi} = 0.4$ in radian, under training overhead $K = 100$. The simulation results are provided in Fig.~\ref{imperfect_star_ris_to_BS}, where $\epsilon_1 = \sqrt{0.9}$ and $\eta_1 = \sqrt{0.5}$. As expected, the localization performance degrades when $\hat{d}$ and $\hat{\varphi}$ increases, especially in the high SNR regime. 

\subsection{Effect of MPCs}
\begin{figure}[t]
	\centering
\includegraphics[width=0.95\linewidth]{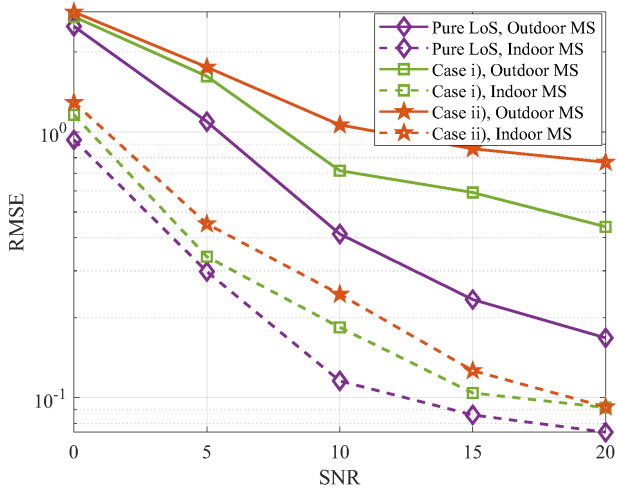}
	\caption{The effect of MPCs on 3D localization with different pairs of $\hat{d}$ and $\hat{\varphi}$ with $K = 100$, $\epsilon_1 = \sqrt{0.9}$, and $\eta_1 = \sqrt{0.5}$. }
		\label{Effect_of_MPCs}
\end{figure}

The proposed localization algorithm purely relies channel parameters associated with the LoS path. Therefore, the strength of the non-line-of-sight (NLoS) path will negatively affect the localization performance. In principle, the stronger the strength, the worse the localization performance. In this subsection, we pick up different setups on the average sum power of the NLoS paths (revealed by the distance) and evaluate accordingly their negative effect on the localization performance. We introduce two MPCs to $\bh_1$, $\bh_2$, and $\bh_3$ while keeping perfect LoS condition for $\bH_4$. The introduced MPCs for $\bh_i$, for $i = 1,2,3$, comprise the following channel parameters: case i) $\{10d_i, \theta_i+ \pi/6, \phi_i + \pi/6 \}$ and $\{10d_i, \theta_i+ \pi/3, \phi_i + \pi/3 \}$ and case ii) $\{5d_i, \theta_i+ \pi/6, \phi_i + \pi/6 \}$ and $\{5d_i, \theta_i+ \pi/3, \phi_i + \pi/3 \}$. The simulation results are shown in Fig.~\ref{Effect_of_MPCs}, where we observe that the stronger the MPCs, the worse the localization performance. Even though there are works on leveraging NLoS paths for enhancing localization performance~\cite{Witrisal2016, Mendrzik2019}, we will leave such an extension for our future investigation.  
\section{Conclusion and Future Work}\label{sec_conclusions}
In this paper, we studied the fundamental 3D localization performance limits of STAR-RIS-empowered millimeter wave MIMO systems for simultaneously serving one indoor and one outdoor MSs. We presented a practical localization algorithm based on ANM, which approaches the theoretical performance limits characterized by our derived CRLBs. In addition, we thoroughly investigated the effect of training overhead, energy splitting at the STAR-RIS, the power allocation between the two MSs, the STAR-RIS design, the imperfectness of STAR-RIS-to-BS channel, as well as the role of the presence of MPCs on the localization performance of the two MSs, offering some useful insights for future practical implementations. In future works, we will extend the theoretical CRLB analyses to general multipath scenarios and exploit the availability of MPCs for further enhancing the localization performance of the two MSs in  the design of practical localization algorithms.  





\bibliographystyle{IEEEtran}
\bibliography{IEEEabrv,Ref}

\begin{thebibliography}{10}
\providecommand{\url}[1]{#1}
\csname url@samestyle\endcsname
\providecommand{\newblock}{\relax}
\providecommand{\bibinfo}[2]{#2}
\providecommand{\BIBentrySTDinterwordspacing}{\spaceskip=0pt\relax}
\providecommand{\BIBentryALTinterwordstretchfactor}{4}
\providecommand{\BIBentryALTinterwordspacing}{\spaceskip=\fontdimen2\font plus
\BIBentryALTinterwordstretchfactor\fontdimen3\font minus
  \fontdimen4\font\relax}
\providecommand{\BIBforeignlanguage}[2]{{%
\expandafter\ifx\csname l@#1\endcsname\relax
\typeout{** WARNING: IEEEtran.bst: No hyphenation pattern has been}%
\typeout{** loaded for the language `#1'. Using the pattern for}%
\typeout{** the default language instead.}%
\else
\language=\csname l@#1\endcsname
\fi
#2}}
\providecommand{\BIBdecl}{\relax}
\BIBdecl

\bibitem{huang2019reconfigurable}
C.~Huang, A.~Zappone, G.~C. Alexandropoulos, M.~Debbah, and C.~Yuen,
  ``Reconfigurable intelligent surfaces for energy efficiency in wireless
  communication,'' \emph{IEEE Trans. Wireless Commun.}, vol.~18, no.~8, pp.
  4157--4170, Aug. 2019.

\bibitem{di_renzo_smart_2019}
M.~{Di Renzo \textit{et al.}}, ``Smart radio environments empowered by
  reconfigurable {AI} meta-surfaces: an idea whose time has come,''
  \emph{{EURASIP} J. Wireless Comm. and Networking}, vol. 2019, p. 129, 2019.

\bibitem{WavePropTCCN}
G.~C. Alexandropoulos, G.~Lerosey, M.~Debbah, and M.~Fink, ``Reconfigurable
  intelligent surfaces and metamaterials: {T}he potential of wave propagation
  control for {6G} wireless communications,'' \emph{IEEE ComSoc TCCN
  Newslett.}, vol.~6, no.~1, pp. 25--37, Jun. 2020.

\bibitem{risTUTORIAL2020}
Q.~Wu, S.~Zhang, B.~Zheng, C.~You, and R.~Zhang, ``Intelligent reflecting
  surface aided wireless communications: {A} tutorial,'' \emph{IEEE Trans.
  Commun.}, vol.~69, no.~5, pp. 3313--3351, May 2021.

\bibitem{RISE6G_COMMAG}
E.~Calvanese~Strinati, G.~C. Alexandropoulos, H.~Wymeersch, B.~Denis,
  V.~Sciancalepore, R.~D'Errico, A.~Clemente, D.-T. Phan-Huy, E.~D. Carvalho,
  and P.~Popovski, ``Reconfigurable, intelligent, and sustainable wireless
  environments for {6G} smart connectivity,'' \emph{IEEE Commun. Mag.},
  vol.~59, no.~10, pp. 99--105, Oct. 2021.

\bibitem{wymeersch2019radio}
H.~{Wymeersch}, J.~{He}, B.~{Denis}, A.~{Clemente}, and M.~{Juntti}, ``Radio
  localization and mapping with reconfigurable intelligent surfaces:
  Challenges, opportunities, and research directions,'' \emph{{IEEE} Veh.
  Technol. Mag.}, vol.~15, no.~4, pp. 52--61, Dec. 2020.

\bibitem{Jiguang2020}
J.~He, H.~Wymeersch, L.~Kong, O.~Silvén, and M.~Juntti, ``Large intelligent
  surface for positioning in millimeter wave {MIMO} systems,'' in \emph{Proc.
  IEEE VTC-Spring}, 2020, pp. 1--5.

\bibitem{Elzanaty2021}
A.~Elzanaty, A.~Guerra, F.~Guidi, and M.-S. Alouini, ``Reconfigurable
  intelligent surfaces for localization: Position and orientation error
  bounds,'' \emph{{IEEE} Trans. Signal Process.}, vol.~69, Aug. 2021.

\bibitem{Alexandropoulos2022}
G.~C. Alexandropoulos, I.~Vinieratou, and H.~Wymeersch, ``Localization via
  multiple reconfigurable intelligent surfaces equipped with single receive
  {RF} chains,'' \emph{{IEEE} Wireless Commun. Lett.}, vol.~11, no.~5, pp.
  1072--1076, 2022.

\bibitem{he2022simultaneous}
J.~He, A.~Fakhreddine, and G.~C. Alexandropoulos, ``Simultaneous indoor and
  outdoor {3D} localization with {STAR-RIS}-assisted millimeter wave systems,''
  in \emph{proc. IEEE Vehicular Technology Conference (Fall)}, September, 2022,
  pp. 1--6.

\bibitem{Tsinghua_RIS_Tutorial}
M.~Jian, G.~C. Alexandropoulos, E.~Basar, C.~Huang, R.~Liu, Y.~Liu, and
  C.~Yuen, ``Reconfigurable intelligent surfaces for wireless communications:
  {O}verview of hardware designs, channel models, and estimation techniques,''
  \emph{Int. Conv. Netw.}, vol.~3, no.~1, pp. 1--32, Mar. 2022.

\bibitem{alexandg_2021}
G.~C. Alexandropoulos, N.~Shlezinger, and P.~del Hougne, ``Reconfigurable
  intelligent surfaces for rich scattering wireless communications: {R}ecent
  experiments, challenges, and opportunities,'' \emph{IEEE Commun. Mag.},
  vol.~59, no.~6, pp. 28--34, Jun. 2021.

\bibitem{Hris_mag}
\BIBentryALTinterwordspacing
G.~C. Alexandropoulos, N.~Shlezinger, I.~Alamzadeh, M.~F. Imani, H.~Zhang, and
  Y.~C. Eldar, ``Hybrid reconfigurable intelligent metasurfaces: {E}nabling
  simultaneous tunable reflections and sensing for {6G} wireless
  communications,'' 2021. [Online]. Available:
  \url{https://arxiv.org/pdf/2104.04690}
\BIBentrySTDinterwordspacing

\bibitem{Yuanwei2021}
Y.~Liu, X.~Mu, J.~Xu, R.~Schober, Y.~Hao, H.~V. Poor, and L.~Hanzo, ``{STAR}:
  Simultaneous transmission and reflection for 360$^{\circ}$ coverage by
  intelligent surfaces,'' \emph{{IEEE} Wireless Commun.}, vol.~28, no.~6, pp.
  102--109, Dec. 2021.

\bibitem{Chenyu2022}
C.~Wu, C.~You, Y.~Liu, X.~Gu, and Y.~Cai, ``Channel estimation for
  star-ris-aided wireless communication,'' \emph{{IEEE} Commun. Lett.},
  vol.~26, no.~3, pp. 652--656, 2022.

\bibitem{Xinwei2022}
X.~Yue, J.~Xie, Y.~Liu, Z.~Han, R.~Liu, and Z.~Ding, ``Simultaneously
  transmitting and reflecting reconfigurable intelligent surface assisted noma
  networks,'' \emph{{IEEE} Trans. Wireless Commun.}, pp. 1--1, 2022.

\bibitem{AoI2022}
G.~C. Alexandropoulos, M.~Crozzoli, D.-T. Phan-Huy, K.~D. Katsanos,
  H.~Wymeersch, P.~Popovski, P.~Ratajczak, Y.~Bénédic, M.-H. Hamon,
  S.~Herraiz~Gonzalez, R.~D'Errico, and E.~Calvanese~Strinati, ``Smart wireless
  environments enabled by riss: Deployment scenarios and two key challenges,''
  in \emph{Proc. Joint European Conference on Networks and Communications \& 6G
  Summit}, Grenoble, France, Jun. 2022, pp. 1--6.

\bibitem{Akdeniz2014}
M.~R. {Akdeniz}, Y.~{Liu}, M.~K. {Samimi}, S.~{Sun}, S.~{Rangan}, T.~S.
  {Rappaport}, and E.~{Erkip}, ``Millimeter wave channel modeling and cellular
  capacity evaluation,'' \emph{{IEEE} J. Sel. Areas Commun.}, vol.~32, no.~6,
  pp. 1164--1179, Jun. 2014.

\bibitem{Tsai2018}
Y.~{Tsai}, L.~{Zheng}, and X.~{Wang}, ``Millimeter-wave beamformed
  full-dimensional {MIMO} channel estimation based on atomic norm
  minimization,'' \emph{{IEEE} Trans. Commun.}, vol.~66, no.~12, pp.
  6150--6163, Dec. 2018.

\bibitem{Wu2019}
Q.~{Wu} and R.~{Zhang}, ``Beamforming optimization for wireless network aided
  by intelligent reflecting surface with discrete phase shifts,'' \emph{{IEEE}
  Trans. Commun.}, pp. 1--1, 2019.

\bibitem{Shahmansoori2017}
A.~Shahmansoori, G.~E. Garcia, G.~Destino, G.~Seco-Granados, and H.~Wymeersch,
  ``Position and orientation estimation through millimeter-wave {MIMO in 5G}
  systems,'' \emph{{IEEE} Trans. Wireless Commun.}, vol.~17, no.~3, pp.
  1822--1835, 2018.

\bibitem{scharf1993geometry}
L.~L. Scharf and L.~T. McWhorter, ``Geometry of the {Cramer-Rao} bound,''
  \emph{Signal Process.}, vol.~31, no.~3, pp. 301--311, Apr. 1993.

\bibitem{Pakrooh2015}
P.~Pakrooh, A.~Pezeshki, L.~L. Scharf, D.~Cochran, and S.~D. Howard, ``Analysis
  of {Fisher} information and the {Cramér–Rao} bound for nonlinear parameter
  estimation after random compression,'' \emph{{IEEE} Trans. Signal Process.},
  vol.~63, no.~23, pp. 6423--6428, Dec. 2015.

\bibitem{golub2013matrix}
G.~H. Golub and C.~F. Van~Loan, \emph{Matrix computations}.\hskip 1em plus
  0.5em minus 0.4em\relax JHU press, 2013.

\bibitem{he20223d}
J.~He, A.~Fakhreddine, C.~Vanwynsberghe, H.~Wymeersch, and G.~C.
  Alexandropoulos, ``{3D} localization with a single partially-connected
  receiving {RIS}: Positioning error analysis and algorithmic design,''
  \emph{arXiv preprint arXiv:2212.02088}, 2022.

\bibitem{Sung2010}
H.~Sung, S.-H. Park, K.-J. Lee, and I.~Lee, ``Linear precoder designs for
  k-user interference channels,'' \emph{{IEEE} Trans. Wireless Commun.},
  vol.~9, no.~1, pp. 291--301, 2010.

\bibitem{he2020anm}
J.~{He}, H.~{Wymeersch}, and M.~{Juntti}, ``Channel estimation for {RIS}-aided
  {mmWave MIMO} systems via atomic norm minimization,'' \emph{{IEEE} Trans.
  Wireless Commun.}, pp. 1--1, 2021.

\bibitem{Yang2016}
Z.~{Yang} and L.~{Xie}, ``Exact joint sparse frequency recovery via
  optimization methods,'' \emph{{IEEE} Trans. Signal Process.}, vol.~64,
  no.~19, pp. 5145--5157, Oct 2016.

\bibitem{Tang2013}
G.~{Tang}, B.~N. {Bhaskar}, P.~{Shah}, and B.~{Recht}, ``Compressed sensing off
  the grid,'' \emph{{IEEE} Trans. Inf. Theory}, vol.~59, no.~11, pp.
  7465--7490, 2013.

\bibitem{Witrisal2016}
K.~Witrisal, P.~Meissner, E.~Leitinger, Y.~Shen, C.~Gustafson, F.~Tufvesson,
  K.~Haneda, D.~Dardari, A.~F. Molisch, A.~Conti, and M.~Z. Win,
  ``High-accuracy localization for assisted living: {5G} systems will turn
  multipath channels from foe to friend,'' \emph{{IEEE} Signal Process. Mag.},
  vol.~33, no.~2, pp. 59--70, 2016.

\bibitem{Mendrzik2019}
R.~Mendrzik, H.~Wymeersch, G.~Bauch, and Z.~Abu-Shaban, ``Harnessing {NLOS}
  components for position and orientation estimation in {5G} millimeter wave
  {MIMO},'' \emph{{IEEE} Trans. Wireless Commun.}, vol.~18, no.~1, pp. 93--107,
  2019.

\end{thebibliography}

\vfill\pagebreak



\end{document}